\documentclass[twocolumn,superscriptaddress,amsmath,amssymb]{revtex4}
\pdfoutput=1
\usepackage{graphicx}
\usepackage{dcolumn,xcolor,ulem}
\usepackage{amsmath} 
\usepackage{amssymb}
\usepackage{amsfonts}
\usepackage{bm}
\usepackage[latin1]{inputenc}

\begin{document}



\title{Potential ``ways of thinking'' about the shear-banding phenomenon\\}

\author{M.A.~Fardin}
\affiliation{Laboratoire Mati\`ere et Syst\`emes Complexes, CNRS UMR 7057\\ Universit\'e Paris Diderot, 10 rue Alice Domont et L\'eonie Duquet, 75205 Paris C\'edex 13, France} 
\affiliation{Department of Mechanical Engineering\\ Massachusetts Institute of Technology, 77 Massachusetts Avenue, MA 02139-4307 Cambridge, USA}
\author{T.J.~Ober}
\affiliation{Department of Mechanical Engineering\\ Massachusetts Institute of Technology, 77 Massachusetts Avenue, MA 02139-4307 Cambridge, USA}
\author{C.~Gay}
\author{G.~Gr\'egoire}
\affiliation{Laboratoire Mati\`ere et Syst\`emes Complexes, CNRS UMR 7057\\ Universit\'e Paris Diderot, 10 rue Alice Domont et L\'eonie Duquet, 75205 Paris C\'edex 13, France} 
\author{G.H.~McKinley}
\affiliation{Department of Mechanical Engineering\\ Massachusetts Institute of Technology, 77 Massachusetts Avenue, MA 02139-4307 Cambridge, USA}
\author{S.~Lerouge}
\altaffiliation[Corresponding author ]{}
\email{sandra.lerouge@univ-paris-diderot.fr}
\affiliation{Laboratoire Mati\`ere et Syst\`emes Complexes, CNRS UMR 7057\\ Universit\'e Paris Diderot, 10 rue Alice Domont et L\'eonie Duquet, 75205 Paris C\'edex 13, France}

\date{\today}

\begin{abstract}
Shear-banding is a curious but ubiquitous phenomenon occurring in soft matter. The phenomenological similarities between the shear-banding transition and phase transitions has pushed some researchers to adopt a `thermodynamical' approach, in opposition to the more classical `mechanical' approach to fluid flows. In this heuristic review, we describe why the apparent dichotomy between those approaches has slowly faded away over the years. To support our discussion, we give an overview of different interpretations of a single equation, the diffusive Johnson-Segalman (dJS) equation, in the context of shear-banding. We restrict ourselves to dJS, but we show that the equation can be written in various equivalent forms usually associated with opposite approaches. We first review briefly the origin of the dJS model and its initial rheological interpretation in the context of shear-banding. Then we describe the analogy between dJS and reaction-diffusion equations. In the case of anisotropic diffusion, we show how the dJS governing equations for steady shear flow are analogous to the equations of the dynamics of a particle in a quartic potential. Going beyond the existing literature, we then draw on the Lagrangian formalism to describe how the boundary conditions can have a key impact on the banding state. Finally, we reinterpret the dJS equation again and we show that a rigorous effective free energy can be constructed, in the spirit of early thermodynamic interpretations or in terms of more recent approaches exploiting the language of irreversible thermodynamics. 
\end{abstract}

\maketitle

By `shear-banding', scientists can mean several things. In the field of material science, `shear-banding' refers to the notion of strain localisation. When a solid material is deformed, the strain can take large values in narrow zones of the sample. Similarly, in the field of soft matter--\textit{i.e.} complex fluids, \textit{i.e.} non-Newtonian fluid mechanics--`shear-banding' refers to the notion of strain rate localisation. When a fluid material is sheared, the strain rate can take large values in narrow zones of the sample. In both case, for solids or for fluids, shear-banding is linked to a sharp inhomogeneity in the deformation or deformation rate field. Clear domains of different strains or strain rates are identifiable. This phenomenology is associated with `complex materials' as it is clearly distinct from the simpler homogeneous deformation or deformation rate fields in ideal Hookean solids, Newtonian fluids, or weakly plastic or shear thinning materials~\cite{Larson99}.\\ 
Even if we restrain ourselves to fluids, the phenomenon has been observed in a variety of systems. Recent experiments include shear-banding in telechelic polymers~\cite{Manneville07,Sprakel08}, emulsions~\cite{Coussot02,Becu06}, dispersions~\cite{Divoux10}, granular materials~\cite{Losert00} or even foams~\cite{Gilbreth06}. There is still no universal framework to describe shear-banding in all of those various systems and there is still no consensus on the mesoscopic mechanisms involved in shear-banding across systems. A persistent idea is that there exists some kind of underlying `flow-structure' coupling. Mechanisms for such coupling include entanglement effects, breakage, liquid-crystalline effects, changes in charge and association,  or changes in topology~\cite{Olmsted08}. Shear-banding may even be a generic macroscopic phenomenon, able to spring out of many different underlying mechanisms. For instance, it was thought for some time that shear-banding was necessarily associated with non-monotonic constitutive relations~\cite{Hunter83,McLeish86}, but it has subsequently been realized that other mechanisms such as stress inhomogeneity inherent to the geometry--in large gap Taylor-Couette geometry for instance--or boundary effects can lead to shear-banding~\cite{consensus}.\\
In this paper, we describe the steady shear-banding phenomenon in micellar solutions~\cite{Berret05,Cates06,Lerouge09}. In the concentration and temperature ranges where micellar solutions exhibit shear-banding, the micelles are structured in long polymer-like flexible chains, which can break and recombine~\cite{Cates06}. Those so-called `wormlike micelles' entangle in a highly viscoelastic network, Maxwellian in the linear regime, \textit{i.e.} possessing a single relaxation time and elastic modulus~\cite{Larson99}. Their simple linear rheology, and the robustness of the shear-banding states in those materials has elevated them to a status of model systems for the understanding of shear-banding in fluids~\cite{Rehage91}. And many see a solid understanding of shear-banding in micellar fluids as a necessary first step toward a clearer view of the phenomenon in general. Since the seminal study by Rehage and Hoffmann~\cite{Rehage91}, much theoretical and experimental effort has been devoted to gain comprehension on the phenomenon. The most recent review on the subject~\cite{Lerouge09} referenced more than three hundred articles, and it is sometimes challenging to find one's way through such an enormous amount of literature. In particular, from a theoretical perspective, many approaches of shear-banding have been proposed and it is sometimes hard to understand their connections.\\
For wormlike micelles, the original idea of an underlying non-monotonic flow curve is most likely to be relevant to almost all experimental situations investigated so far. Theoretically, this is justified by the strong reductionist rationale provided by the reptation-reaction model~\cite{Cates06}. This statistical theory of micelles is an adaptation of the reptation theory of polymers, including micellar breaking and recombination processes~\cite{Cates06}. This theory can both predict linear rheology with great accuracy and the onset of shear-banding, due to an underlying inhomogeneity of the flow curve~\cite{Cates06}. Nonetheless, this model becomes highly intractable in the non-linear flow regime where shear-banding occur, and it has been of little or no help to understand shear-banding in more details. To reach better tractability in the non-linear flow regime, constitutive models have been used~\cite{Larson99,Cates06,Fielding07}. Constitutive models, which rely on very general material frame indifference principles, do not usually contain all the information on the microstructure dynamics and deal with coarse-grained quantities defined at the macroscopic scale. They usually at least include a tensorial stress field, and a tensorial velocity gradient field~\cite{Larson99}. \\
In the last few years, one particular constitutive model has been used extensively, namely, the diffusive Johnson-Segalman (dJS) equation and its mechanistic interpretation~\cite{Cates06,Fielding07}. It is most likely one of the simplest tensorial model able to predict shear banding. Because it is a quasi-linear model~\cite{DPL} with only few parameters, it is analytically tractable in many cases. Thus, notwithstanding some known shortfalls--especially its violation of the Lodge-Meissner relationship, and its troubles in extensional flows and step strains~\cite{Larson88}--the dJS model has generally been a very useful rational guide to interpret empirical data and to initiate new experiments. Under the impulse of Olmsted, Lu, Radulescu, Fielding and others~\cite{Fielding07}, it has been used to predict the onset of shear-banding in various flow geometries~\cite{Radulescu99,Lu00,Radulescu00}, to discuss transient effects~\cite{Radulescu03} and most importantly, to realize that shear-banding flows could themselves become unstable to elastic instabilities~\cite{Fielding05,Fielding07b,Fielding10}.  \\
It is clearly an achievement to be able to explain many different aspects of shear-banding flows using a single model. But, what do we really mean by ``using a single model''? If different authors have used the same equation, they have not necessarily interpreted the equation in a single way. To use Feynman's words, there are different ``ways of thinking'' about the shear-banding phenomenon using the dJS equation~\cite{Feynman83}. We think that when assessing the success of the dJS model, it is crucial to distinguish its syntactic power--\textit{i.e.} linked to its mathematical structure--from its various physical interpretations. This point is crucial to better understand the distinction that has often been made between a `mechanical' and a `thermodynamical' approach~\cite{Berret94,Schmitt95,Porte97}. Traditionally, the dJS model is thought of as a `mechanical' approach. Nonetheless, Fielding has already shown that it is possible to construct models including ingredients from both perspectives~\cite{Fielding03}. The so-called dJS-$\phi$ model includes a `mechanical subspace' and a `concentration subspace', coupled to each others~\cite{Fielding03}. More broadly, Cates and Fielding have recently remarked that the distinction between `mechanical' banding instabilities and shear-induced structural instabilities is likely to be less clear-cut than was once thought~\cite{Cates06}. Later, Olmsted added that in practice there is little difference between the two~\cite{Olmsted08}. This claim was made even clearer last year by Sato \textit{et al.}~\cite{Sato10}. In their paper, they derived a pseudo-thermodynamic potential from the dJS model. Unfortunately, we think that this achievement is tempered by the fact that they use a reduction of the number of degrees of freedom, obtaining an equation that is not strictly equivalent to the original dJS model.\\
In this heuristic review, we describe why the apparent opposition between those approaches has slowly faded away over the years. To support our discussion, we give an almost historical account of different interpretations of a single equation, the diffusive Johnson-Segalman (dJS) equation, in the context of shear-banding. We restrict ourselves to dJS, but we show that the equation can be written in various equivalent forms usually associated with opposite approaches. We first review briefly the origin of the dJS model and its initial rheological interpretation in the context of shear-banding. Then we describe the analogy between dJS and reaction-diffusion equations. In the case of anisotropic diffusion, we show how the dJS governing equations for steady flow are analogous to the equations of the dynamics of a particle in a quartic potential. Going beyond the existing literature, we then draw on the Lagrangian formalism to describe how the boundary conditions can have a key impact on the banding state. Finally, we show for the first time without any algebraic simplification that a rigorous effective free energy can be constructed from the dJS model, in the spirit of early thermodynamic interpretations~\cite{Berret94,Schmitt95,Dhont99,Porte97} or of more recent approaches exploiting the language of irreversible thermodynamics~\cite{Bautista02,Manero07,Bautista07}.\\
This paper is not an exhaustive review of the many approaches to shear-banding and their various achievements. More detailed reviews already exist on the matter~\cite{Cates06,Fielding07}. We believe that instead, this paper can provide a translation scheme between different approaches. Along the way this translation effort allows us to derive some interesting new features of shear-banding flows.\\

\begin{figure}
\begin{center}
\includegraphics[trim = 5mm 0mm 5mm 0mm,width=7cm,clip]{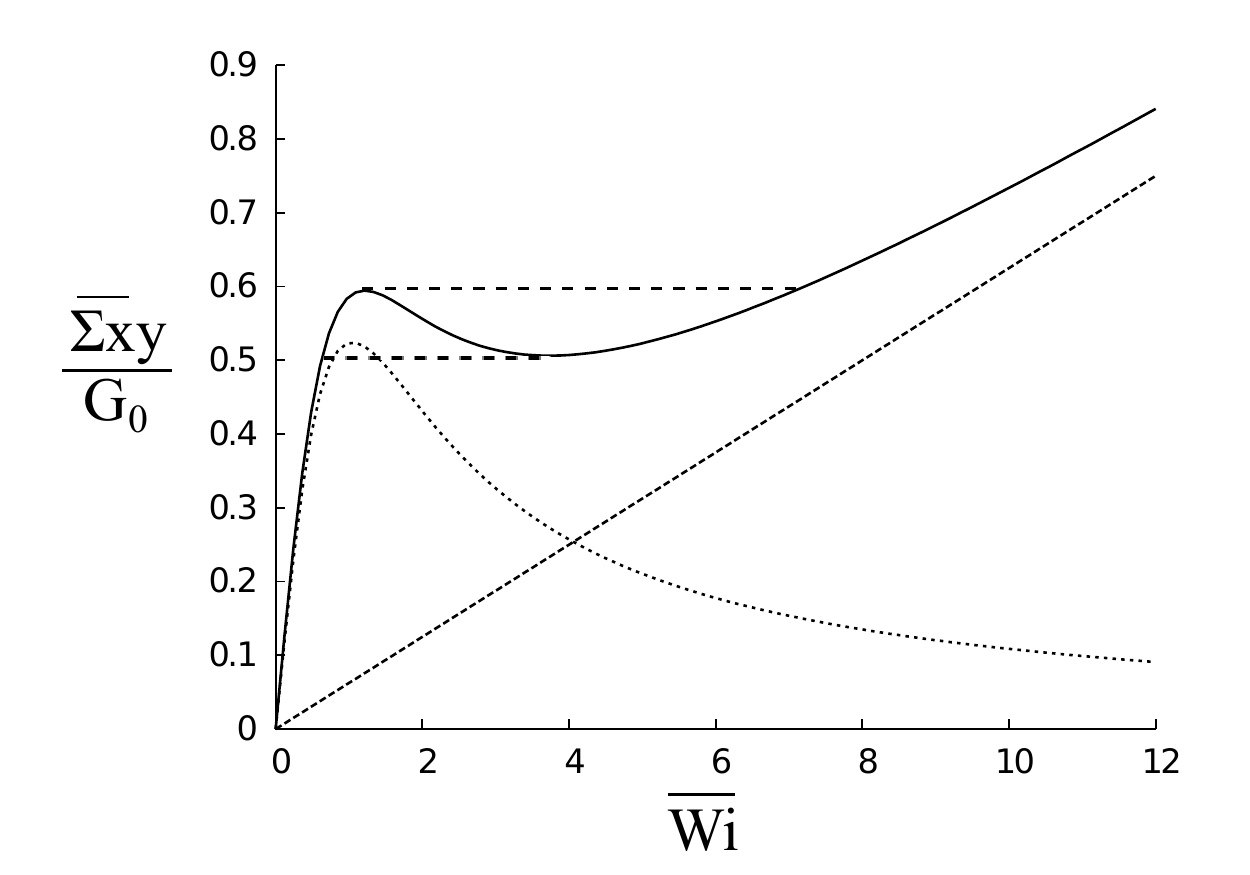}
\caption{Non-monotonic homogeneous flow curve for simple steady shear of the JS model, for $a=0.3$ and $\eta=1/16$. $\overline{Wi}\equiv \frac{U\lambda}{d}$ is the global Weissenberg number. The fine dashed lines are the polymeric and solvent contributions to the shear stress. The horizontal dashed lines highlight the range of stress in which the flow curve is multivalued. Without any `plateau selection rule', the stress can take any value in between the two horizontal dashed lines.   
\label{JScurve}}
\end{center}
\end{figure}

\section{JS model and sketch of shear-banding}
\label{JSandSketch}
The model originally proposed by Johnson and Segalman \cite{Johnson77} was a modification to the Upper Convected Maxwell model (UCM), the canonical rheological model for viscoelasticity~\cite{Larson99,DPL}. In order to allow rate-dependent material properties in steady simple shear, the JS model supposes that network strands in the material can slip with respect to a purely affine deformation. This slipping yields an effective velocity gradient field which is given by
\begin{equation}
{\bf{L}}\equiv\nabla \vec v - 2\zeta \bf{D}
\end{equation}
\noindent where $\zeta$ is a scalar slip coefficient in the range $0\leq\zeta\leq1$, $\nabla \vec v$ is the velocity gradient tensor, and ${\bf{D}}\equiv\frac{1}{2}((\nabla \vec v)^t+\nabla \vec v)$ is the strain rate tensor. The new convected derivative operator is 
\begin{equation}
\widehat{(~)}\equiv\frac{D(~)}{Dt}-{\bf{L}}^t\cdot(~)-(~)\cdot{\bf{L}}
\label{Gordon}
\end{equation}
\noindent where $\frac{D(~)}{Dt} \equiv \frac{\partial (~)}{\partial t}+\vec v\cdot \nabla (~)$ is the material derivative. The resulting constitutive equation for the polymeric stress tensor $\bf{T}$, is written
\begin{equation}
{\bf{T}}+\lambda \widehat{\bf{T}}=2\eta_p \bf{D}
\label{T_gov1}
\end{equation}
\noindent where $\lambda$ is the `polymer' relaxation time and $\eta_p\equiv G_0 \lambda$ is the `polymer' viscosity, defined with respect to the `polymer' elastic modulus.  \\
In recent publications~\cite{Fielding07}, Eq.~(\ref{T_gov1}) is usually rewritten in an equivalent form involving a rescaled `slip parameter' $a\equiv 1-2\zeta$, where $-1\leq a\leq1$: 
\begin{equation}
\frac{D{\bf{T}}}{Dt}+\frac{{\bf{T}}}{\lambda}=a\Big({\bf{D}}\cdot{\bf{T}} + {\bf{T}}\cdot{\bf{D}}\Big)+\Big({\bf{\Omega}}\cdot {\bf{T}} - {\bf{T}}\cdot {\bf{\Omega}}\Big)+2G_0{\bf{D}}
\label{T_gov2}
\end{equation}
\noindent where ${\bf{\Omega}}\equiv\frac{1}{2}((\nabla \vec v)^t-\nabla \vec v)$ is the vorticity tensor. For the case of $a=1$, we obtain the UCM model, if $a=-1$ we have the lower convected Maxwell model, and if $a=0$ we have the co-rotational Maxwell model~\cite{DPL}. In a steady simple shear flow, we have $\vec v=[u(y), 0, 0]$, and gradients in the flow properties exist only in the $y$-direction, between $y=0$ and $y=d$. We choose a reference frame in which the plate at $y=0$ is fixed, \textit{i.e.} $u(0)=0$, while the plate at $y=d$ is moving with $u(d)=U$. Then, we can define the characteristic global shear rate as $\bar{\dot\gamma}\equiv \frac{U}{d}$, itself frame independent. The homogeneous solution for this flow is defined as the solution in which the local shear rate is constant $\dot\gamma(y)=\bar{\dot\gamma}$~\cite{homogeneous}. Then, if $|a|\neq 1$, the polymeric shear stress, obtained from eq. (\ref{T_gov2}) and shown in Fig.~\ref{JScurve}, is non monotonic, and drops to zero. To cure this pathology, the common practice is to add a `solvent' contribution to the stress. Then, the total deviatoric stress of the material is given by the sum of the polymeric stress and the solvent stress ${\bf{\Sigma}}\equiv {\bf{T}} + 2\eta_s{\bf{D}}$. In a simple shear geometry, the momentum balance imposes the value of the local deviatoric shear stress to be constant $\Sigma_{xy}(y)=\overline{\Sigma_{xy}}$, where $\Sigma_{ij}$ are the components of $\bf{\Sigma}$. Then, as pictured in Fig. ~\ref{JScurve}, the global flow curve $\overline{\Sigma_{xy}}=f(\bar{\dot\gamma})$ is known to be non-monotonic if $\eta\equiv\frac{\eta_s}{\eta_p}<\frac{1}{8}$~\cite{Kolkka88,Espanol96,Cates06,Fielding07}, which was identified early as being one of the sufficient criteria for triggering shear-banding~\cite{Hunter83,McLeish86}. Motivated by early studies on the JS model~\cite{Renardy87,Kolkka88,Renardy95}, by empirical observations~\cite{Rehage91} and by exploiting an analogy with first order phase transitions~\cite{Berret94}--especially the pressure/specific volume graphs--it was realized that this non-monotonic flow curve was the signature of an instability of the homogeneous flow. Quickly, it was understood that in a range of global shear rates $[\bar{\dot\gamma}_1 , \bar{\dot\gamma}_2]$, in the vicinity of the decreasing part of the flow curve, the flow would become inhomogeneous, $i.e.$ $\dot\gamma(y)\neq\bar{\dot\gamma}$. For global shear rates `in the shear-banding regime', $i.e.$ for $\bar{\dot\gamma}_1<\bar{\dot\gamma}<\bar{\dot\gamma}_2$, the flow would be split in domains with local shear rates $\bar{\dot\gamma}_1$ and $\bar{\dot\gamma}_2$, with the proportion of the sample in high ($\alpha$) or low ($1-\alpha$) shear rates domains defined as following a `lever rule':
\begin{equation}
\bar{\dot\gamma}=\alpha \bar{\dot\gamma}_2 + (1-\alpha) \bar{\dot\gamma}_1
\label{leverrule}
\end{equation}
From the momentum balance, the total stress must stay homogeneous in simple shear. Thus, during the entire shear-banding regime, increasing the global shear rate will not induce any global shear stress increase, just change the relative proportions of the two bands. The additional injected power is used in turning more of the sample into the high shear rate domain, $i.e.$ in increasing $\alpha$. But in contrary to first order phase transitions where the Maxwell equal area law~\cite{Berkeley5} gives a criterion to select precise values of $[\bar{\dot\gamma}_1 , \bar{\dot\gamma}_2]$ and thus also the `plateau' value of the stress, it was unclear what criterion could be used in the context of the JS model. For some time, it was even believed that such `plateau selection' would be doomed in any mechanical approach such as JS~\cite{Porte97}.  

\section{dJS and plateau selection}
\label{dJSandPlateau}
After the first few studies of shear-banding using the JS model, it was quickly realized that some key ingredient was missing. The degeneracy in the selection of the plateau value was linked to the absence in the JS model of a characteristic length scale that would set the interface thickness between the shear bands, a point raised earlier in consideration of a simpler but similar model~\cite{Spenley96}. Subsequently, many arguments, inspired from dynamical systems~\cite{Lu00} or from kinetic considerations~\cite{Radulescu99,Radulescu03} were proposed to rationalize this new length scale as arising from a diffusion term--or `non-local term'--missing in the JS equation. Rheological equations such as JS or UCM can be derived from the kinetic theory of dumbells ~\cite{Larson99,DPL}. And indeed, a careful treatment of the Fokker Planck equation underlying the kinetic theory leading to the JS models--or even the UCM model--brings a diffusion term coming from the finite size of the dumbells~\cite{ElKareh89}. From those considerations, the JS model was modified to account for this diffusion term, leading to the so-called diffusive JS model (dJS). If we allow for anisotropic stress diffusion, the diffusion term takes the form $\bf{\nabla} \cdot\bf{{\cal D}} \cdot \bf{\nabla} {\bf{T}}$~\cite{Bird07}. Then, we can define the dJS model as:
\begin{align}
\frac{D{\bf{T}}}{Dt}+\frac{{\bf{T}}}{\lambda}=a\Big({\bf{D}}\cdot{\bf{T}} + {\bf{T}}\cdot{\bf{D}}\Big)+\Big({\bf{\Omega}}\cdot {\bf{T}} - {\bf{T}}\cdot {\bf{\Omega}}\Big)\nonumber\\
+2G_0{\bf{D}}+ \bf{\nabla} \cdot\bf{{\cal D}} \cdot \bf{\nabla} {\bf{T}}
\label{T_gov_diffiso}
\end{align}
\noindent Evidently, the units of the diffusion coefficients ${\cal{D}}_{ij}$ are [m$^2$/s]. From each diffusion coefficient we can define a diffusion length scale $\ell_{ij}\equiv \sqrt{{\cal D}_{ij}\lambda}$, which will be involved in the scaling of the typical width of the interface between shear bands.\\
The additional governing equations for the isothermal and incompressible flow are the continuity equation and the Cauchy momentum equation:
\begin{align}
\nabla \cdot \vec v &= 0 \\
\rho\Bigg(\frac{\partial}{\partial t} + \vec v \cdot \nabla\Bigg)\vec v &= \nabla \cdot\bigg({\bf{T}}+2\eta_s{\bf{D}}-p{\bf{I}}\bigg)
\label{Cauchy}
\end{align}
$p$ is the isotropic pressure and $\rho$ is the density of the fluid, including the polymeric and the solvent part. {\bf{I}} is the unit tensor.  

\section{Simple shear and dimensionless groups}
\label{SimpleShear}
In simple shear flow, the governing equations for the polymer stress take the form given by:
\begin{align}
\label{simpleshear}
&\tau_{xx}+\lambda\frac{\partial}{\partial t} \tau_{xx} - (1 + a) \lambda\dot\gamma \tau_{xy} = \ell_{xx}^2 \frac{\partial^2}{\partial y^2} \tau_{xx}\\
&\tau_{yy}+\lambda\frac{\partial}{\partial t} \tau_{yy} + (1 - a) \lambda\dot\gamma \tau_{xy} = \ell_{yy}^2 \frac{\partial^2}{\partial y^2} \tau_{yy}\\
&\tau_{zz}+\lambda\frac{\partial}{\partial t} \tau_{zz}  =\ell_{zz}^2 \frac{\partial^2}{\partial y^2} \tau_{zz}\\
&\tau_{xy}+\lambda\frac{\partial}{\partial t} \tau_{xy} +\frac{1}{2}\lambda\dot\gamma \bigg[ (1 - a) \tau_{xx} - (1 + a)\tau_{yy} \bigg] \nonumber\\
&= \eta_p\dot\gamma + \ell_{xy}^2 \frac{\partial^2}{\partial y^2} \tau_{xy}
\label{simpleshear2}
\end{align}
\noindent Finally, we take the $x$-component of the momentum equation Eq.~(\ref{Cauchy}), 
\begin{equation}
\rho\frac{\partial u}{\partial t} = \frac{\partial \tau_{xy}}{\partial y} + \eta_s\frac{\partial^2 u}{\partial y^2}
\label{Cauchyx}
\end{equation}
Eq.~(\ref{simpleshear}-\ref{Cauchyx}) are the governing equations of the fluid dynamics in simple shear flow. Those equations are still dimensional, and seem to involve many quantities (variables and parameters). Dimensional analysis suggests dimensionless groups that reduce the apparent number of quantities~\cite{Buckingham14}. Basic dimensional analysis can reduce the number of quantities by three, because there are only three fundamental dimensions involved. But actually, we can recast the different variables and parameters in six categories, using stress, time, viscosity, length and density as independent units. Then, from there we can easily construct the relevant dimensionless groups. The stress variables turn into $\tau_{xx}/G_0$, $\tau_{yy}/G_0$, $\tau_{zz}/G_0$, $\tau_{xy}/G_0$. The time variable turns into the inverse of the Deborah number $De^{-1} \equiv t^* \equiv t/\lambda$ and we introduce the local Weissenberg number $Wi\equiv \lambda\dot\gamma(y) = \lambda \frac{\partial u}{\partial y}$~\cite{Dealy10}. We have already defined the viscosity ratio $\eta\equiv\eta_s/\eta_p$. The length variable turns into $y^*=y/d$ and we introduce Knudsen numbers for stress diffusion $\xi_{ij}\equiv \ell_{ij}/d$. From the density, we could construct the usual Reynolds number, but since we are mainly interested in the creeping flow regime, we instead use the elasticity number $\mathcal{E }\equiv \frac{\overline{Wi}}{\overline{Re}}\equiv \frac{\lambda^2 G_0}{\rho d^2}$. $\overline{Wi}\equiv \frac{U\lambda}{d}$ is the global Weissenberg number, and $\overline{Re}\equiv\frac{\rho d U}{\eta_p}$ is the global Reynolds number. Conventionally, the Reynolds number is constructed from the total viscosity of the fluid $\eta_p +\eta_s$, but in effect we are interested in cases where $\eta_p \gg \eta_s$ and so this does not really matter for our purpose. Finally, we also need to retain the slip parameter $a$ (already dimensionless). 

\section{Reaction-diffusion interpretation}
\label{ReactDiffInt}
To simplify even more the form of the governing equations, we can use dimensionless variables introduced previously by Radulescu \textit{et al.}~\cite{Radulescu99}, $K \equiv \sqrt{1-a^2} Wi$, $S \equiv \sqrt{1-a^2} \frac{\tau_{xy}}{G_0}$, $N \equiv (1 - a) \frac{\tau_{xx}}{2G_0} - (1 + a) \frac{\tau_{yy}}{2G_0}$ and $Z \equiv (1 - a) \frac{\tau_{xx}}{2G_0} + (1 + a) \frac{\tau_{yy}}{2G_0}$. The total dimensionless stress is then $\sigma\equiv S+\eta K$ and is equal to its global value $\bar{\sigma}$ everywhere in the sample, for a steady simple shear flow. Moreover, Radulescu \textit{et al.} used a constant Knudsen number $\xi_0$ for every stress component. Then, after ignoring the z-component, we can transform the governing equations~(\ref{simpleshear}-\ref{Cauchyx}) into their dimensionless counterparts:
\begin{align}
\label{N_gov}
&\frac{\partial}{\partial t^*}N = K S - N + \xi_0^2 \frac{\partial^2}{\partial y^{*2}} N\\
\label{Z_gov}
&\frac{\partial}{\partial t^*}Z = -Z + \xi_0^2 \frac{\partial^2}{\partial y^{*2}} Z \\
\label{S_gov}
&\frac{\partial}{\partial t^*}S = -K N + K - S + \xi_0^2 \frac{\partial^2}{\partial y^{*2}} S \\
\label{K_gov}
&\frac{1}{\mathcal{E}}\frac{\partial}{\partial t^*} K = \frac{\partial^2}{\partial y^{*2}}\bigg(S + \eta K\bigg)
\end{align}
Where the last equation is obtained by differentiating~\ref{Cauchyx} with respect to \textit{y}. The utility of those equations is their independence from the value of the slip parameter, $a$, when $|a|\neq 1$~\cite{Radulescu99,Radulescu00}. When the problem was expressed in this form, its connection to the more general class of reaction-diffusion problems was noticed by Radulescu \textit{et al.}~\cite{Radulescu99}. The governing equation for $Z$ being decoupled, its analysis is not usually carried out~\cite{Radulescu99,Radulescu00}. Moreover, in most experimental situations we have $\mathcal{E}\gg 1$. Thus, from Eq. (\ref{K_gov}), it is apparent that the dynamics of $K$ happen on a much shorter time scale than the dynamics of $S$ and $N$. For this reason, the evolution of the kinematics with time is not seen as an independent dynamical variable in the limit $\mathcal{E}\to \infty$, and the reaction-diffusion problem is written in terms of only two degrees of freedom, a dimensionless shear stress $S$ and a dimensionless normal stress difference $N$~\cite{Radulescu99,Radulescu00}:
\begin{align}
\label{ReactDiff}
& \partial_t \binom{S}{N} = \xi_0^2 \partial^2_{y^*} \binom{S}{N} + \mathbb{C}(S,N;K),\\
&\noindent \text{where} \nonumber\\
& \mathbb{C}(S,N;K) \equiv  \binom{-S+K-KN}{-N+KS} \nonumber
\end{align}
Within this framework, Radulescu \textit{et al.} derived a variety of important properties of shear banding flows~\cite{Radulescu99,Radulescu00}. We will reassess those properties in a different approach in the following, but we will make repeated connections with the approach of Radulescu \textit{et al.}.  

\section{Steady simple shear and the particle analogy}
\label{PartAna}
\begin{figure}
\begin{center}
\includegraphics[width=7cm,clip]{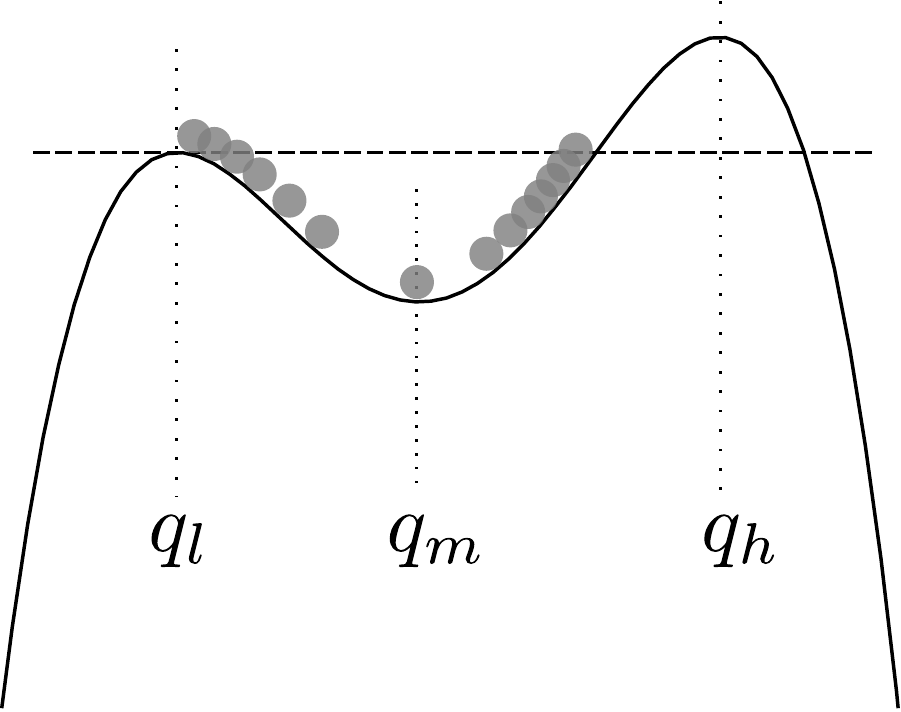}
\caption{Sketch of the quartic potential with regions of fast moving and slow moving particle. $q_l <q_m <q_h$ are the equilibrium points. The dashed line is the total energy line above which the motion is unbounded, because the particle may escape the local potential well from the left.   
\label{potshape}}
\end{center}
\end{figure}
In the following, we consider steady flows, \textit{i.e.} $De=0$~\cite{Dealy10}. Transient effects have already been discussed~\cite{Radulescu99,Radulescu03}, and an update is on the way~\cite{Radulescu11}. For steady flows, Eq. (\ref{ReactDiff}) gives 
\begin{equation}
\xi_0^2 \partial^2_{y^*} \binom{S}{N} = -\mathbb{C}(S,N;K)
\label{ReactDiff_stdy}
\end{equation}
Radulescu \textit{et al.}~\cite{Radulescu99} noted that this equation is analogous to Newton's second law for the movement of a fictitious particle. The same analogy had been used previously by Spenley \textit{et al.}~\cite{Spenley96}, also in the context of shear-banding, and more generally by Pomeau~\cite{Pomeau86}. But with two degrees of freedom $S$ and $N$, the complexity of Eq. (\ref{ReactDiff_stdy}) prevented Radulescu \textit{et al.} from seeking an analytic solution to the inhomogeneous shear-banding flow. \\
If the diffusion in the non-linear flow regime becomes anisotropic, we cannot use a single Knudsen number $\xi_0$. If diffusion only concerns the shear component of the stress, then $\xi_{ij}=0$ for every $i$ and $j$ except $\xi_{xy}=\xi$. In this case, dimensional considerations would suggest that the diffusion can be expressed in terms of the shear rate tensor, rather than the stress tensor. In this alternate version of dJS, the diffusion term is of the form $2\eta_p {\cal D}_0 \nabla^2{\bf{D}}$, with ${\cal D}_0 \equiv \frac{\xi^2 d^2}{\lambda}$. Note that the sign of this term has to be taken in accordance with the definition of the sign of the shear rate. This type of term was used recently by Sato \textit{et al.}~\cite{Sato10}, because it is much more mathematically tractable, as was already apparent in a simpler diffusive model used by Dhont~\cite{Dhont99}. In the following, we will assume the diffusion to be of this anisotropic kind. The differences with taking an isotropic stress diffusion as in Radulescu \textit{et al.}~\cite{Radulescu99,Radulescu00} are discussed in the appendices of Sato \textit{et al.}~\cite{Sato10}. \\
With diffusion on the shear rate, and for steady flow ($De=0$), the system of Eqs. (\ref{N_gov}-\ref{K_gov}) is replaced by 
\begin{align}
\label{sdy_N_gov}
&N = K S  \\
\label{sdy_Z_gov}
&Z = 0 \\
\label{sdy_S_gov}
&\xi^2 \frac{\partial^2}{\partial y^{*2}} K  = -K N + K - S \\
\label{sdy_K_gov}
&\frac{\partial^2}{\partial y^{*2}}\bigg(S + \eta K\bigg)=0 
\end{align}
Eq.~(\ref{sdy_K_gov}) is redundant with the homogeneity of the stress discussed above. Eq.~(\ref{sdy_Z_gov}) is trivial. We can use Eq.~(\ref{sdy_N_gov}) to replace $N$ in Eq.~(\ref{sdy_S_gov}). Then, it is convenient to regard $K$ as the main variable of our problem, and by using $\sigma=\bar{\sigma}= S+\eta K$, we express $S$ as a function of $K$ and we reach
\begin{align}
& K\big(1-K(\bar\sigma-\eta K)\big)-(\bar\sigma-\eta K) - \xi^2 \frac{\partial^2 K}{\partial y^{*2}} = 0 \nonumber \\
\Leftrightarrow\quad & \xi^2 \frac{\partial^2 K}{\partial y^{*2}} = \eta K^3 - \bar\sigma K^2 + (1+\eta)K - \bar\sigma
\label{kappa_2nd_ODE}
\end{align}
If $y^*$ is reinterpreted as a time variable, then Eq.~(\ref{kappa_2nd_ODE}) is analogous to the equation of the motion of a particle of mass $m\leftrightarrow \xi^2$ and position $q\leftrightarrow K$ under a force deriving from a quartic potential energy function $V(q)$ depending on the parameters $\bar\sigma$ and $\eta$. This analogy was stated explicitly in the recent study by Sato \textit{et al.}~\cite{Sato10}, building on the idea invoked earlier with Eq. (\ref{ReactDiff_stdy})~\cite{Radulescu99}. Using the chain rule, we can transform Eq. (\ref{kappa_2nd_ODE}) into its `energy form':
\begin{align}
\label{kappa_energy_eq}
&\frac{1}{2}m \dot q^2 + V(q)=E  \\
 V(q) & \equiv F q^4 +D q^3 + C q^2 + B q  + A\nonumber\\
         & \equiv -\frac{1}{4}\eta q^4 + \frac{1}{3}\bar\sigma q^3 - \frac{1}{2}(1+\eta)q^2 + \bar\sigma q
\end{align}
Where $E$ is the total energy of the system, which is conserved. And we have used $\dot q \leftrightarrow \frac{d K}{dy^*}$. To make the analogy explicit, we drop our previous usage of $t$ and use $t\leftrightarrow y^*$. By using `$\leftrightarrow$' or `stands for' to define the quantities in the particle analogy, we want to differentiate these definitions with regular definitions using `$\equiv$'. When using `$\leftrightarrow$', the new quantities are linked with a new interpretation of the syntactic object. In the language of mathematicians, we `define a new model' for the dJS equation~\cite{Hodges97}. This is what we mean by `making an analogy'. In some sense, we could have even used $\leftrightarrow$ in defining the dimensionless variables $K$, $S$, $N$ and $Z$, since they carried the new meaning given in the framework of reaction-diffusion by Radulescu \textit{et al.}~\cite{Radulescu99}. \\
We will see that this new interpretation helps us realize the full syntactic power of the dJS equation, but before exploiting further the analogy, from Eq. (\ref{kappa_2nd_ODE}), we can already obtain the non-monotonic homogeneous flow curve we mentioned previously. If we force the flow to be homogeneous, $K(y)=\bar{K}\equiv (1-a^2)^{1/2} \bar{Wi}$. Then $\partial_{y*^2} K \equiv 0$ (where we use $\partial_{y*^2}$ to stand for $\frac{\partial^2}{\partial y*^2}$) and Eq. (\ref{kappa_2nd_ODE}) reaches
\begin{align}
\eta \bar{K}^3 - \bar\sigma \bar{K}^2 + (1+\eta)\bar{K} - \bar\sigma & = 0\nonumber  \\
\bar\sigma &=\frac{\bar{K}}{1+\bar{K}^2} +\eta \bar{K} 
\end{align}
Note that this solution can also be obtained by taking $\xi^2=0$, \textit{i.e.} the particle has no mass. This is another way of thinking about why the inhomogeneous flow curve can only be obtained by the addition of a diffusive term in JS. 

\section{Properties of the potential}
\label{PotProp}
In the particle analogy, it is important to realize that regions where the shear rate changes abruptly correspond to time intervals where the particle is moving fast, i.e. near local minima of the potential $V(q)$--the `inner solution' described by Radulescu et al.\cite{Radulescu99,Radulescu00}. On the contrary, regions of fairly constant shear rate, \textit{i.e.} linear velocity profile, correspond to time intervals where the particle is moving slowly, near the `turning points' of the potential $V(q)$~\cite{Landau76}--the `outer solution'~\cite{Radulescu99,Radulescu00}. The potential $V(q)$ is a quartic potential. The general solution of the motion of a particle in a quartic potential is well known and involves elliptic functions for $q(t)$. A short recent summary of important analytical results can be found in Sanchez \textit{et al.}~\cite{Sanchez93}. In particular, Sanchez \textit{et al.}~\cite{Sanchez93} describe the various possible shapes for the quartic potential, depending on relations between the coefficients of the potential ($A,B,C,D,F$). Evidently, the shape of the potential is critical in determining the solution. To simplify the study of the shape of the potential, we follow Sanchez \textit{et al.} in the following two steps. Firstly, we define the three equilibrium points, roots of the algebraic equation $\frac{dV(q)}{dq}=0$. Lets name the equilibrium locations following the ordering $q_l < q_m < q_h$. Those solutions are real if $\eta<1/8$, \textit{i.e.} when the homogeneous flow curve is non-monotonic.  Since $F<0$, $\lim_{q\to \pm \infty} V(q) =-\infty$, and by continuity, we already know that $q_l$ and $q_h$ are unstable equilibrium points whereas $q_m$ is a stable equilibrium, as illustrated on Fig.~\ref{potshape}. The particle will move fast near $q_m$ and slowly near $q_l$ and $q_h$. \\
To be able to use tabulated coefficient relations~\cite{Sanchez93}, it is convenient to rescale our variable $q$ to eliminate the linear term of the potential.  We can make use of the middle root $q_m$ of the equation $\frac{dV(q)}{dq}=0$. The idea is to translate the coordinate system such that it is centred on the middle root value. We introduce the variable $x\equiv q-q_m$. Then, from Sanchez \textit{et al.}~\cite{Sanchez93} we can introduce a new potential $V^*(x)\equiv A^* x^2 + B^* x^3 +C^* x^4$, such that 
\begin{equation}
\text{Eq. (\ref{kappa_energy_eq})} \Longleftrightarrow \frac{1}{2}m\dot x^2 + V^*(x)=E^*=E-V(q_m) 
\label{x_energy_eq}
\end{equation}
 With, 
\begin{equation} 
   \begin{cases}
 A^*\equiv C+3 D q_m +6 F q_m^2 = -\frac{\eta+1}{2} +\bar{\sigma}q_m -(3/2)\eta q_m ^2\\
 B^*\equiv D+4 F q_m = \bar{\sigma}/3 -\eta q_m\\
 C^*\equiv F = -\eta /4
  \end{cases}
\label{coefrelations}
\end{equation}
\begin{figure}
\begin{center}
\includegraphics[width=7.8cm,clip]{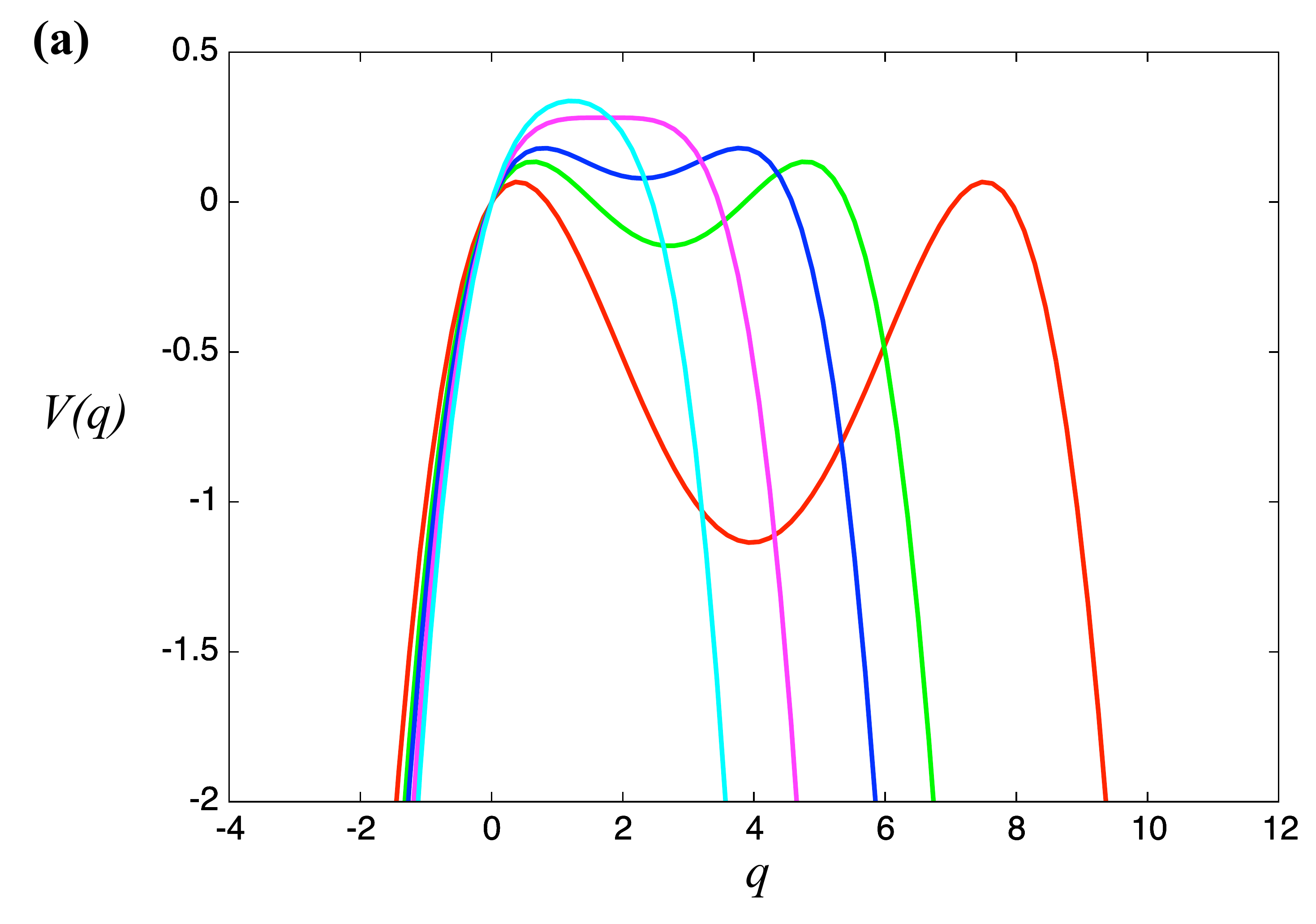}
\includegraphics[width=7.8cm,clip]{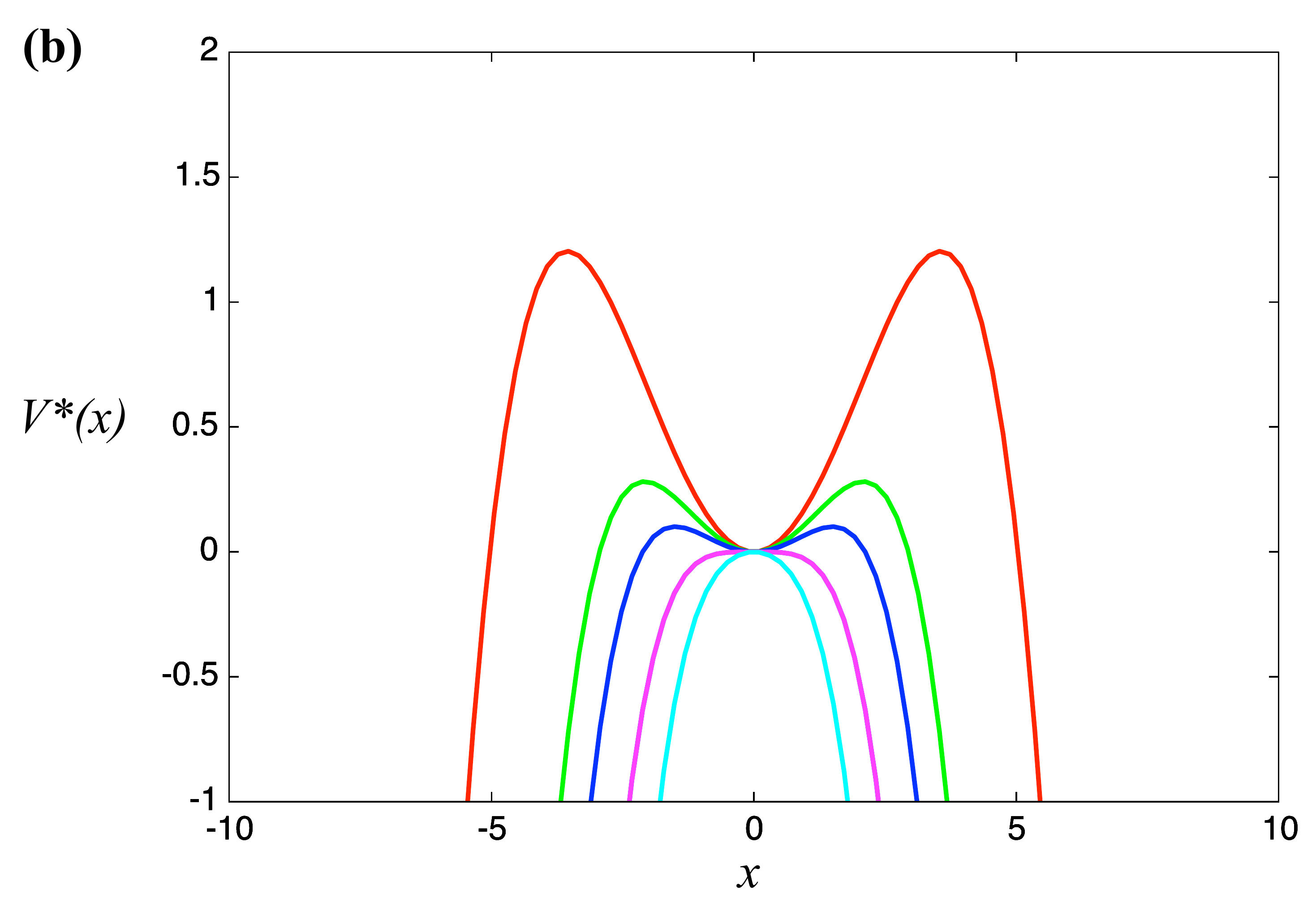}
\caption{Plot of (a) the original quartic potential $V(q)$ and (b) the rescaled potential $V^*(x)$, for $\eta=0.03,0.06,0.08,1/8,0.2$. 
\label{Potentials}}
\end{center}
\end{figure}
In the rheological framework, we know that the stress has to be homogeneous. Nonetheless, we anticipate the local shear rate to be inhomogeneous. The shear rate associated with a given stress can take a range of values. On the global scale this leads to the so-called `stress-plateau'. For a value of $\bar\sigma$ one can realize several values of $\bar K$. This degeneracy means that given a value of stress in the bulk $\bar\sigma$, there is no `bulk mechanism' capable of selecting a value of $\bar K$. As it will become clearer in section~\ref{Sato_unconstrained}, this means that the solution $q(t)$ contains an arbitrary parameter linked to the arbitrariness of the value of $\bar K$. As a particular consequence, the equilibrium points $q_l$ and $q_h$ need to have the same relative stability, \textit{i.e.}, as stated in Sato \textit{et al.} the potential $V(q)$ or $V^*(x)$ needs to be symmetric~\cite{Sato10}. \\
For the potential to be symmetric, the coefficients must be related by the relation ${B^*}^2=4A^*C^*$~\cite{Sanchez93}. Using Eqs. (\ref{coefrelations}), this equation, together with $\frac{dV}{dq}\Big\vert_{q_m}=0$ reads
\begin{align}
\label{fundeq1}
& {B^*}^2=4A^*C^* \Leftrightarrow \quad \frac{\bar\sigma^2}{9} -\frac{\eta^2 q_m^2}{2}+\frac{\eta\bar\sigma q_m}{3}-\frac{\eta(\eta+1)}{2}=0\\
&\frac{dV}{dq} \Big\vert_{q_m}=0 \quad \Leftrightarrow\quad  \eta q_m^3 - \bar\sigma q_m^2 + (1+\eta)q_m - \bar\sigma  = 0
\label{fundeq2}
\end{align}
We therefore have a system of two algebraic equations with unknowns $(q_m,\bar\sigma)$. From those two equations, we can eliminate the variable $q_m$ to reach
\begin{align}
\bar\sigma (9\eta-18\eta^2-2\bar\sigma^2)=0 \nonumber \\
\Rightarrow \quad \bar\sigma =\pm \frac{3\sqrt{\eta-2\eta^2}}{\sqrt{2}} \quad \text{or} \quad \bar\sigma=0
\label{sigmaeq}
\end{align}
The negative and positive values correspond to the arbitrariness of the stress sign. Thereafter, we keep $\bar\sigma>0$ by convention. From Eq. (\ref{sigmaeq}) it is evident that for a given value of the material parameter $\eta$, there is a unique value of stress $\bar\sigma$ corresponding to the non homogeneous solution. \\
If we seek the complete solution to the inhomogeneous flow, we should then replace any instance of the mean stress $\bar\sigma$ by its value depending on $\eta$. In particular, we can now find the expression for the three equilibrium points of the potential:
\begin{equation}
   \begin{cases}
& q_{l} = \frac{\sqrt{1/\eta - 2} - \sqrt{1/\eta -8}}{\sqrt{2}} \\
& q_m=\frac{\sqrt{1/\eta -2}}{\sqrt{2}} \\
& q_{h} = \frac{\sqrt{1/\eta - 2} + \sqrt{1/\eta -8}}{\sqrt{2}} 
 \end{cases}
\label{Keq}
\end{equation}
Note that since the potential is symmetric, $q_m=\frac{q_l + q_h}{2}=\frac{\bar\sigma}{3\eta}$.\\
From Eq. (\ref{sigmaeq},\ref{Keq}), we can rewrite $A^*$, $B^*$ and $C^*$ as functions of $\eta$ only,
\begin{equation} 
   \begin{cases}
 A^*= \frac{1-8\eta}{4}\\
 B^*= 0\\
 C^*= -\eta /4
  \end{cases}
\label{newcoefrelations}
\end{equation}
Fig. ~\ref{Potentials} draws the original potential $V(q)$ and the rescaled potential $V^*(x)$, for a few values of $\eta$.  As already mentioned, the potential $V(q)$ as a unique equilibrium point $q_m$ when $\eta \geq 1/8$. At the point $\eta=1/8$, $q_m$ is a multiple root and $A^*=0$. We call this point, the `critical point'.\\
The characteristic frequency of the particle near $q_m$ is given by the harmonic approximation of $V(x)$ near $x=0$, $i.e.$ by $\omega_0=\sqrt{\frac{1-8\eta}{4m}}$. $1/\omega_0$ is the time scale of the problem. This is a crucial point, as the time interval $[0,1]$ was indefinite and we now recognize that the `time' for the particle to translate from $q_l$ to $q_h$ is in units of the harmonic period near $q_m$. Note that at the critical point, $\omega_0 = 0$, because the lowest order of the potential near $q_m$ is quartic,  the particle time scale diverges.\\
Returning to the original rheological framework, we can identify the inverse of the harmonic frequency with the dimensionless width of the interface between domains of different shear rates, $w \leftrightarrow 1/\omega_0$. This correspond to the `inner solution' in the reaction-diffusion framework~\cite{Radulescu00}. Thus, the interface width appears to diverge as $\eta$ approaches the critical point~\cite{notediv}.

\section{Naive flow curve and law of equal distances}
\label{NaiveFC}
\begin{figure}
\begin{center}
\includegraphics[trim = 31mm 0mm 0mm 0mm, width=7cm,clip]{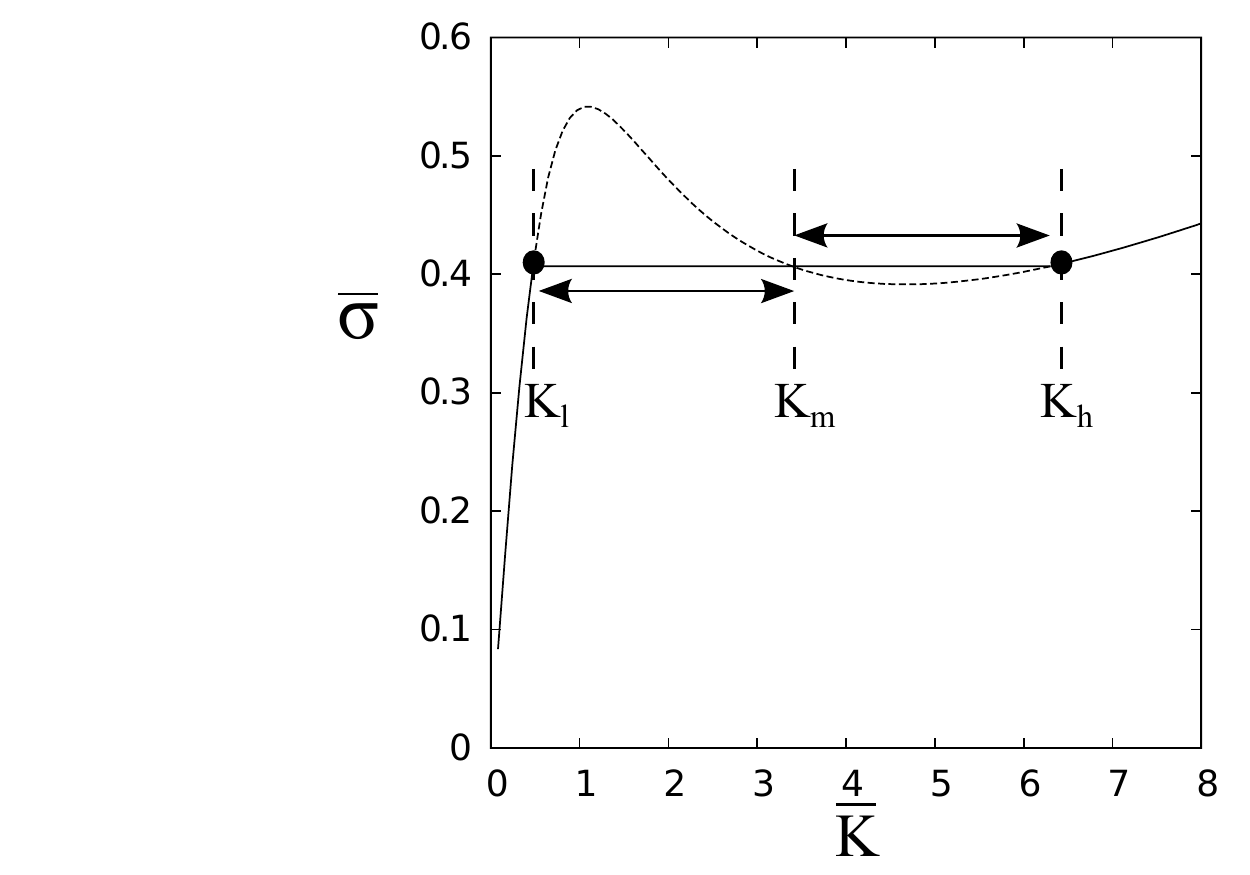}
\caption{Dimensionless homogeneous and inhomogeneous flow curve for $\eta=0.04$. The dashed line is the unstable homogeneous solution. The black circles highlight the connection points $K_l$ and $K_h$ between the inhomogeneous and homogeneous solution. The connection points are related by $K_m - K_l =K_h-K_m$. 
\label{JSmidrule}}
\end{center}
\end{figure}

Returning to the rheological framework, the values $K_l\leftrightarrow q_l$ and $K_h \leftrightarrow q_h$ are the outer solutions. As explained already in \cite{Radulescu99,Radulescu00}, in the limit where $\xi\ll 1$, one has to seek matching between the homogeneous solution with the shear-banding solution at $K_l$ and $K_h$.  Then, $K_l$ and $K_h$ are identified as the boundary of the stress plateau with magnitude $\bar\sigma$. Fig. \ref{JSmidrule} displays the homogeneous and inhomogeneous flow curves for $\eta=0.04$, highlighting the connections points $K_l$ and $K_h$ between the two solutions. Ultimately, we would expect the flow curve measurable for a steady simple shear to read,
\begin{equation}
\bar{\sigma}(\bar{K}) =
  \begin{cases}
   \frac{\bar K}{1+\bar K^2} + \eta\bar K & \quad \text{if}\quad \eta \geq \frac{1}{8} \quad\text{or}\quad \bar{K}\notin [K_l , K_h]\\
     \frac{3\sqrt{\eta-2\eta^2}}{\sqrt{2}}      &\quad \text{if } \quad\eta < \frac{1}{8} \quad\text{and}\quad \bar{K}\in [K_l , K_h]
  \end{cases}
\label{totalFCsmallxi}
\end{equation} 
The plateau intersects the homogeneous flow curve at three locations $K_l$, $K_m$ and $K_h$. Then, the symmetry of the potential implies $K_m=\frac{K_l + K_h}{2}$. Thus the plateau is such that $K_m - K_l =K_h-K_m$. By analogy with the law of equal areas, we can call this the `law of equal distances'. A similar law was found by Dhont on a scalar model analogue to dJS~\cite{Dhont99}. This would be an easy criterion, but it does not seem to be recovered by numerical simulations of the dJS model~\cite{Fielding05,Fielding07}. We have forgotten a key ingredient in our derivation. This key ingredient was also missing in Sato's derivations~\cite{Sato10}.

\section{Sato's constrained solution}
\label{Sato_unconstrained}
In Sato \textit{et al.}, the authors make use of the naive flow curve to obtain a solution for the shear rate profile. Just considering the order of Eq. (\ref{kappa_2nd_ODE}), they consider two boundary conditions for $K$, $\partial_{y^*} K=0$ at $y^*=0$ and $1$, assumed to come from $\partial_{y^*} S\mid_{0,1}=0$, which is supposed to represent the fact that there is no flux of polymeric components at the wall~\cite{Sato10,Rossi06}. In the particle analogy, the boundary conditions translate into the requirement that the particle starts with no initial velocity at $t=0$ and ends with no velocity at $t=1$ (in units of $1/\omega_0$.) If we require that the outer part of the inhomogeneous solutions connect with the homogeneous solutions~\cite{Radulescu99,Radulescu00}, then we assume that $q(0)=q_l$ and $q(1)=q_h$ and $\dot q(0)=0$ and $\dot q(1)=0$, the particle is moving `from left to right' on the potential. Note that by symmetry we could have chosen $q(0)=q_h$ and $q(1)=q_l$ as well, \textit{i.e.} the particle moves from right to left. Since the total energy of the particle is conserved, its value is given by the value of the potential at $q_l$ or $q_h$. Then, in the reduced variable $x$, the energy of the system is given by,
\begin{equation}
E^* = V^*(q_l-q_m)=\frac{(8\eta-1)^2}{16\eta} 
\label{energy}
\end{equation}
We can then return to Eq. (\ref{x_energy_eq}), which reads,
\begin{equation}
\frac{1}{2}m\dot x^2 +\Big(\frac{1-8\eta}{4}\Big) x^2 -\frac{\eta}{4} x^4 = \frac{(8\eta-1)^2}{16\eta} 
\label{x_final_eq}
\end{equation}
This equation is separable, $[E^* -V^*(x)]^{-1/2} dx = \sqrt{\frac{2}{m}}dt$. Using our convention that the particle moves from left to right, we can obtain the solution, written with the particle notation $x(t)$, or with the dimensionless rheological notation $K(y^*)$,
\begin{align}
 x(t) &= \frac{q_h-q_l}{2} \tanh\Big[\omega_0 (t-t_0)\Big] \nonumber \\
 \Leftrightarrow q(t) &= \frac{q_h+q_l}{2}+\frac{q_h-q_l}{2} \tanh\Big[\omega_0 (t-t_0)\Big]\nonumber \\
 \leftrightarrow K(y^*) &= \frac{K_h+K_l}{2}+\frac{K_h-K_l}{2} \tanh\Big[\frac{y^*-y^*_0}{w}\Big]\nonumber\\
  K(y^*) &= K_m+\frac{\Delta K}{2} \tanh\Big[\frac{y^*-y^*_0}{w}\Big]
 \label{solutionsmallxi}
\end{align} 
where we have defined the plateau range $\Delta K\equiv K_h-K_l$. At this point, $y^*_0$ is an unknown constant which represents the location of the interface between bands in the gap. Following the naive shear-banding scenario described in the introduction, from Eq. (\ref{leverrule}), we expect that $y^*_0=(1-\alpha)=\frac{K_h-\bar{K}}{K_h -K_l}$. Indeed, it is the value set in Sato \textit{et al.}~\cite{Sato10}. But as we will see in section \ref{leverrulemodif}, this value is not strictly rigorous. 

\section{Modification of the lever rule by non-local effects}
\label{leverrulemodif}
The rigorous value of $y^*_0$ needs to be deduced from the requirement that the integral of the shear rate in the sample should be equal to its macroscopic value. This was indeed noted by Sato \textit{et al.} and stated in dimensionless form (eq. (7) of~\cite{Sato10}):
\begin{equation}
\int_0^1 K(y^*) dy^*=\bar{K}
\label{constraint}
\end{equation}
By integrating the profile obtained in Eq. (\ref{solutionsmallxi}) we can obtain a rigorous expression for $y^*_0$,
\begin{equation}
y^*_0 = \frac{w}{2} \log\Big[{\frac{e^{\frac{1}{w}} - e^{\frac{2\alpha - 1}{w}}} {e^{\frac{2\alpha - 1}{w}} - e^{-\frac{1}{w}}  } }\Big]
\label{intpos}
\end{equation} 
\begin{figure}
\begin{center}
\includegraphics[ width=7cm,clip]{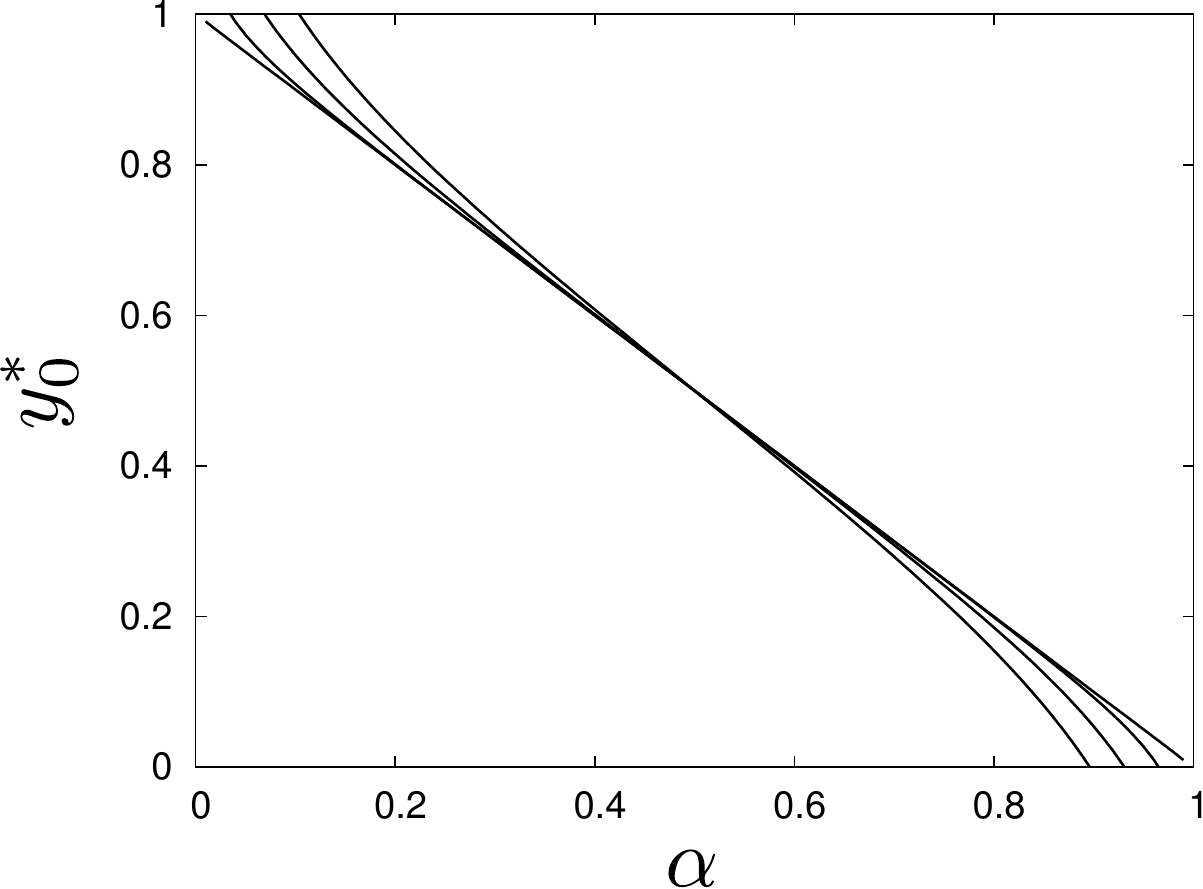}
\caption{Modification of the lever rule $y^*_0 (\alpha)$ for various values of $w=0.01, 0.1, 0.2, 0.3$. To the best of our knowledge, the dJS model had been used only once to study non-local effects, in a numerical simulation never published but communicated in a conference in 2008~\cite{Dambrine08}. In this respect, Eq. (\ref{intpos}) is the first analytical result showing the modification of the lever rule by non-local effects. Nonetheless, this analytic result has only a heuristic purpose since the precise incorporation of the constraint on the average value of the shear rate--described in section~\ref{Lagrangiansec}--would modify the form of this result. 
\label{leverbckdown}}
\end{center}
\end{figure}
Then, only at the lowest order in $(2\alpha - 1)$ and $1/w$, we find $y^*_0 = 1-\alpha$, \textit{i.e.} the interface is located at the position expected from the lever rule. If the interface is too close to the walls, and/or if the width of the interface $w$ starts to be large, this approximation is invalid. This was explicitly stated in the reaction-diffusion framework by Radulescu \textit{et al.}~\cite{Radulescu99}, in terms of ``non-asymptotic effects''. The case where $w$ becomes large can be induced if the Knudsen number $\xi$ is not negligible. It corresponds to the limit where ``non-local effects'' are important~\cite{Masselon08,Masselon10}.  Fig.~\ref{leverbckdown} shows the modification of the lever rule $y^*_0 (\alpha)$ by non-local effects for various values of $w$. A key point is that if $w\neq 0$, there are always non-local effects when one of the shear bands--low or high--is small. In this limit, our assumption that $K(0)=K_l$, $K(1)=K_h$, $\partial_{y^*} K(0)=0$ and $\partial_{y^*} K(1)=0$ is inconsistent, because the width of the interface is such that the shear rate cannot reach its asymptotic values $K_h$ and $K_l$ in the gap interval $y^* \in [0,1]$. As stated in Radulescu \textit{et al.}, matching rigorously happens only if  $y^* \in [-\infty,\infty]$~\cite{Radulescu99}. Moreover, the errors we seem to have made on our boundary conditions are related. For instance, if the interface is close to the wall at $y^*=0$, then $K(0)=K_l$ and $\partial_{y^*} K(0)=0$ will be poorly satisfied whereas $K(1)=K_h$ and $\partial_{y^*} K(1)=0$ will be approximately correct. Where does this pathology come from?
   
\section{Lagrangian formalism and holonomic constraint }
\label{Lagrangiansec}
In the particle analogy, it seems that we could not impose $q(0)=q_l$ and $q(1)=q_h$ and $\dot q(0)=0$ and $\dot q(1)=0$.  In the analogy, Eq. (\ref{constraint}) correspond to a constraint on the average position of the particle during the time interval $[0,1]$. Actually, if one exploits the analogy to its full potential, one realizes that the conditions we had imposed on the motion of our particle were violating causality. If the particle starts at the equilibrium location $q(0)=q_l$, with no initial velocity, there is no mechanism--it seems--that would control the particle dynamics such that it ends at $q(1)=q_h$ with no velocity. We are missing the cause of the motion and trying to impose it after the fact, enforcing Eq. (\ref{constraint}). This anomaly comes from the fact that the integral equation (\ref{constraint}) has to be understood as a holonomic constraint~\cite{Goldstein02}, \textit{i.e.} a constraint on the degrees of freedom of the particle. This constraint reduces the number of boundary conditions we can impose. Generally, the integral constraint has to be understood as providing an additional force that can `push on the particle' in a way that would satisfy the requirement on the time average of the position. This fact was not incorporated in the study by Sato \textit{et al.}~\cite{Sato10}. The inclusion of the holonomic constraint can be done in a consistent way by expressing the particle dynamics in the Lagrangian formalism using the principle of least action~\cite{Landau76,Goldstein02},  
\begin{align}
\delta \mathcal{S}_c(q,\dot{q}, t)\equiv \delta \Big\{\int_0^1{\mathcal{L}_c(q,\dot q,t) dt}\Big\} =0 \nonumber \\
\mathcal{G}(q)\equiv \Big\{\int_0^1{q~dt}\Big\} - \bar{q}=0
\label{optimization}
\end{align}
where $\mathcal{S}_c$ and $\mathcal{L}_c (q,\dot q,t)\equiv T - V \equiv \frac{1}{2} m \dot q^2 - V(q)$ are the action and Lagrangian constrained by $\mathcal{G}$. To solve for Eqs.~(\ref{optimization}) we can use the method of Lagrange multipliers~\cite{Goldstein02}. We introduce a new `unconstrained action' $\mathcal{S}_0$ and a new `unconstrained Lagrangian' $\mathcal{L}_0$ taking the constraint into account through the addition of a Lagrange multiplier $h$.
\begin{equation}
\text{Eqs.(\ref{optimization})} \Leftrightarrow \delta \mathcal{S}_0(q,\dot{q}, t)\equiv \delta \Big\{\int_0^1{\mathcal{L}_0(q,\dot q,t) dt}\Big\} =0
\label{unconstrainedoptimization}
\end{equation}
With $\mathcal{L}_0(q,\dot q,t) \equiv \mathcal{L}_c(q,\dot q,t) + h q$. From the least action principle on the unconstrained action, we can obtain the unconstrained equivalent of Eq. (\ref{kappa_2nd_ODE}) or Eq. (\ref{kappa_energy_eq}) using the Euler-Lagrange equation~\cite{Landau76,Goldstein02}. We then solve the unconstrained Euler-Lagrange equation to get $q(t)$ up to a constant determined by satisfying the constraint~\cite{Goldstein02}. The unconstrained equivalent of Eq.(\ref{kappa_energy_eq}) is
\begin{equation}
\frac{1}{2}m \dot q^2 + V(q)- h q=E_u 
\label{unconstr_energy_eq}
\end{equation}
Where $E_u$ is the unconstrained value of the energy, fully taking into account the integral constraint. From here, the procedure we described previously regarding the naive unconstrained solution would be similar. Since it goes beyond the scope of this review, the full solution to the unconstrained problem will be given elsewhere~\cite{Fardin11}, but we can understand already the impact of imposing the integral constraint in a consistent way.

\section{Modification of the flow curve}
\label{ModFC}
The first effect of the holonomic constraint is the introduction of the Lagrange multiplier that provides the additional force necessary to recover causality in the motion of the particle, by modifying the relative stability of the equilibrium points $q_l$ and $q_h$. Typically, if we keep the same convention for the direction of motion of the particle (from left to right), we expect $h$ to be close to zero if $\bar q\simeq q_l$ and largest when $\bar q\simeq q_h$. So typically, $h$ will be a monotonic function of $\bar q$. And then, since $\bar\sigma$ and $q_l$, $q_m$ and $q_h$ are now functions of $h$, it means that they are also functions of $\bar q$. In the rheological framework, this means that the value of the stress plateau and of what we thought of before as the plateau boundaries $K_l$ and $K_h$ are actually changing as $\bar K$ varies. Typically, since $h=0$ if $\bar K=K_l$, we expect the naive value of the plateau given in Eq. (\ref{sigmaeq}) to be the value of the stress for $\bar K=K_l$, and $K_l$ given by Eq. (\ref{Keq}) to be the actual value of the local shear rate in the sample. But as $\bar K$ increases to values larger than $K_l$, then $h\neq0$ and both the stress and $K_h$ can change. When taken together with the fact that simulations are usually conducted with a fully isotropic stress diffusion--${\cal D}_0 \nabla^2{\bf{T}}$-- the effect of the Lagrange multiplier could explain why the flow curve obtained in simple steady shear flow by numerically solving the dJS model do not follow the naive law of equal distances~\cite{Fielding05,Fielding07}. Also, the local value of shear rate in the high shear rate band can change slightly as the proportion of the high shear rate band increases, \textit{i.e.} as $\bar K$ increases. This has been noticed recently in experiments~\cite{Lettinga09,Feindel10}, and it would be interesting to see if numerical simulations capture this behaviour as well. To the best of our knowledge, a plot giving the local value of $K_h$ in the high shear rate band, as a function of $\bar K$ has never been drawn out of simulations for the dJS model. 

\section{``Competition between shear-banding and wall slip''}
\label{Compet}
\begin{figure}
\begin{center}
\includegraphics[trim = 0mm 0mm 0mm 0mm, width=7cm,clip]{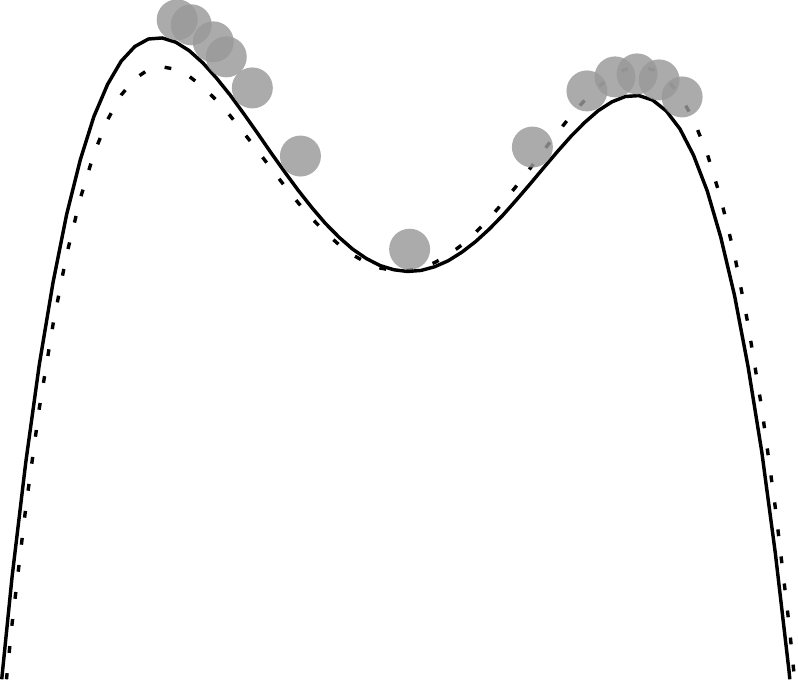}
\caption{Sketch of the dynamics of a particle in a tilted quartic potential. The particle starts with $q(0)=q_l$ and $\dot q(0)=0$, and the final velocity and position at $t=1$ are imposed by the dynamics. 
\label{skewedPot}}
\end{center}
\end{figure}
The second effect of the holonomic constraint is to reduce the number of effective degrees of freedom in the problem. We cannot impose initial and final positions and velocities independently of the value of $\bar q$ \textit{i.e.} of $h$. For instance, if we impose the initial velocity and position, the final position and velocity will be a function of both those initial conditions and $h$, \textit{i.e.} $\bar{q}$. Based on experiments~\cite{Lettinga09} and in order to insure continuity with the homogeneous solution at $\bar K=K_l$,we suggest that a robust boundary condition is $K(0)=K_l$ for any $\bar K$ in the banding regime, \textit{i.e} the shear rate in the low shear rate band is consistently constant. We can then apply one more boundary condition.  If we impose $\partial_{y^*} K(0)=0$, as proposed by Sato et al.~\cite{Sato10}, then we will not be able to control the values of $K(1)$ and $\partial_{y^*} K(1)$. This point can be understood in the particle analogy with the help of Fig.~\ref{skewedPot}. If the particle starts at $q(0)=q_l$, with no velocity, then, for a given value of $\bar q$, the Lagrange multiplier tilts the potential, and at $t=1$, the position and velocity of the particle are imposed by the dynamics. This point will be developed further in a future publication~\cite{Fardin11}, but we believe that this is a strong indication that an apparent slip at the boundary with the high shear rate band may be a genuine feature of shear-banding flows. Because, as can be intuited from Fig.~\ref{skewedPot}, if the potential is tilted, the particle motion near $t=1$ can start accelerating again. In the rheological framework, it means that we can have a thin layer of larger shear rate near $y^*=1$. In general, by lowering the number of boundary conditions we can impose, the holonomic constraint leads to much more freedom on shear-banding states that are possible. \\
In what we just discussed, we took the perspective of fixing two boundary conditions with no slip and then looking at the impact on the other boundary, as a function of $h$, $i.e.$ $\bar q$. But in other experimental situations, the reverse perspective could be just as instructive. Typically, if the material sheared is expected to slip a lot at the moving wall, we can impose slip boundary conditions (on $K(1)$ and on $\partial_{y^*} K(1)$) and then study the resulting state as a function of $h$, $i.e.$ $\bar q$. This perspective would allow the investigation of the effect of large values of the Lagrange multiplier $h$. Remember that in the particle analogy, $h$ has the effect of tilting the potential. For some large values of $h$, the shape of the potential may change dramatically, \textit{i.e.} the number of equilibrium points may go from three ($q_l$, $q_m$ and $q_h$) to only one. In this case, the dynamics of the particle would become unbounded~\cite{Landau76,Goldstein02} and the very assumption of steadiness of the flow would be compromised. This case may be relevant to the dynamics described by Wang \textit{et al.} in terms of ``phase diagrams'' for steady shear-banding or wall slip dominated flows~\cite{Wang11}. \\
The large range of possible shear-banding states was exhibited in a recent numerical study of the dJS model for various boundary conditions~\cite{Adams08}. The necessity of boundary layers was also pointed out in a recent publication exploiting another model of shear-banding~\cite{Cromer11}. Experimentally, the idea of a ``competition between shear-banding and wall slip'' was explicit in the study by Lettinga and Manneville~\cite{Lettinga09}. Generally, the fact that wall slip is usually a genuine but understudied feature of complex fluid flows was expressed recently by R. Buscall~\cite{Buscall10} . Overall, we believe that the Lagrangian interpretation of the particle analogy to the dJS governing equations in simple shear could give the first rigorous analytic rationale for the relations between boundary conditions and shear-banding. The sensitivity of shear-banding states was already mentioned in early studies by Lu, Olmsted and Radulescu, in terms of ``weak-universality''~\cite{Radulescu99,Lu00}, but it is still an area of investigation that needs more theoretical and experimental effort.

\section{Revival of the thermodynamic interpretation}
\label{RevivalThermo}
For readers coming from a more `thermodynamic' rather than `mechanistic' background, it may be apparent already that the Lagrangian formalism for the particle dynamics analogy can readily be re-interpreted (again!) in terms of an effective free energy. In some sense, this effective free energy is reminiscent of early thermodynamical approaches, but stress and deformation are not simple conjugate variables~\cite{Porte97}. From Eq. (\ref{unconstrainedoptimization}) one just needs to map the mechanical variables of the particle analogy, back to their original values. To make the syntax even more transparent, let us use $\phi\leftrightarrow K$ and $\mathcal{F} \leftrightarrow \mathcal{S}_0$. Then instead of interpreting $\delta \mathcal{S}_0=0$ as the least action principle on the particle dynamics, we can interpret $\delta \mathcal{F} =0$ as the minimization of a free energy defined as
\begin{align}
\mathcal{F} \equiv  \int_0^1 dy^* \Big\{\frac{1}{2}\xi^2 \nabla \phi^2 +F \phi^4 + D \phi^3 + C \phi^2 + B \phi \nonumber \\
- h \phi\Big\}
\label{freeenergy}
\end{align}
With the coefficient values given by Eq. (\ref{kappa_energy_eq}) and the external field $h$ whose value facilitates the conservation of the spatially averaged value of the field $\phi$:
\begin{equation}
\int_0^1{dy^* \phi} =\bar\phi
\label{conservedfield}
\end{equation}
This is none other than a conservative `$\phi 4$' model, or Ginzburg-Landau, with an external field $h$~\cite{Goldenfeld92}. This model has been studied in countless frameworks and many derivations from the literature may be readily re-interpreted in the context of shear-banding. The free energy Eq. (\ref{freeenergy}), together with the conservative dynamics imposed by the constraint Eq. (\ref{conservedfield}) belong to the class of `model B', or Cahn-Hilliard dynamics~\cite{Hohenberg77}. The essential point is that the dJS model can be interpreted in a `thermodynamic way' without the need for the algebraic simplifications proposed in Sato \textit{et al.}~\cite{Sato10}--where they reduced the number of degrees of freedom by studying shear-banding infinitesimally close to the critical point at $\eta=1/8$.

\section{Bridging the gap with the BMP model}
\label{BMPmodBridge}
One of the side effects of having a successful canonical model to approach a problem is that this model may overshadow other `concurrent' but promising approaches. The dJS model has been very successful so far, and we have even seen in this paper how far reaching this simple `quasi-linear'~\cite{DPL} model can be. Nonetheless, other approaches have been published and successfully compared to experimental data~\cite{Cates06}. In this last section, we wish to connect to one of those particular models, the Bautista-Manero-Puig model (BMP)~\cite{Bautista02,Manero07,Bautista07}. The BMP approach interprets the shear-banding as explicitly linked to a transition in the mesoscopic structure of the fluid. The idea is to use the UCM model, but with a viscosity locally changing with the strain rate. Together with the UCM mechanical balance expressed as a function of the fluid total viscosity, one constructs an evolution equation for the fluidity $\varphi$--\textit{i.e.} the inverse viscosity--of the material:
\begin{equation}
\frac{\partial\varphi}{\partial t} = \frac{\varphi_0 -\varphi}{\lambda_0} + k_0 (1+\vartheta\dot\gamma) (\varphi_\infty - \varphi) \bf{T:D}
\label{fluidity}
\end{equation}  
where $\varphi_0$ and $\varphi_\infty$ are the fluidity--\textit{i.e.} inverse viscosity--for zero and infinite shear, $\lambda_0$ is the typical relaxation time for the mesoscopic structure to relax to equilibrium, ``$k_0$ is the kinetic parameter at zero-shear strain and the proportionally factor $\vartheta$ is the shear-banding intensity parameter'~\cite{Manero07}. For a steady simple shear flow, the momentum balance and the UCM equation together with Eq. (\ref{fluidity}) reduce to a single equation~\cite{Manero07}:
\begin{equation}
\bar{\sigma} - K + k_0 \lambda_0 (\lambda+\vartheta K)(\varphi_\infty - \frac{K}{\bar\sigma}\varphi_0) \bar{\sigma}^2 G_0^2 \frac{K}{\lambda} =0 
\label{BMPeq}
\end{equation}  
where $\bar{\sigma}$ and $K$ have the values defined previously for $a=1$. The parameters $\lambda$ and $G_0$ are the relaxation time and elastic modulus in the linear regime such that $\eta_0=\varphi_0^{-1}=\lambda G_0$. This is a cubic equation for $K$, in a similar format than Eq. (\ref{kappa_2nd_ODE}) with $\xi=0$.  If we set $\varphi_\infty^{-1}\equiv \lambda_\infty G_0$ and $k\equiv k_0 \lambda_0 G_0$, and with some algebraic manipulations, we can make the syntactic similarity more transparent,
\begin{equation}
\bar{\sigma} - \Big[1-\bar\sigma^2 \frac{k}{\lambda_\infty}\Big] K + \Big[\frac{k}{\lambda}(\frac{\vartheta}{\lambda_\infty}\bar\sigma -1)\bar\sigma\Big] K^2 - \Big[\frac{k\vartheta}{\lambda^2}\bar\sigma\Big]K^3 =0 
\label{BMPeq2}
\end{equation}  
Within the framework of the BMP model Eq. (\ref{BMPeq}) or equivalently Eq. (\ref{BMPeq2}) are seen as having been derived from a quartic potential. Then, the plateau selection also corresponds to the requirement that this potential be symmetric~\cite{Bautista02}. To the best of our knowledge, it was never proven thoroughly that this requirement brings a single value for $\bar\sigma$. Overall, it seems that the BMP and dJS model in simple shear differ only by the values of the coefficients of the potential. Nonetheless--again--this last mapping directly merits some comments. Firstly, Eq. (\ref{BMPeq2}) is generally more complex than Eq. (\ref{kappa_2nd_ODE}) with $\xi=0$, in the sense that it has more parameters ($k$, $\vartheta$, $\lambda_\infty$, etc...). \\
Secondly, the equation is also more complex because it is second order in $\bar\sigma$ rather than linear. This is a key point, since, if we couple Eq.~(\ref{BMPeq2}) with the new symmetry requirement ${B^*}^2=4A^* C^*$--with the new values for the coefficients--then, after eliminating the new equivalent of $q_m$, we would get an equation for $\bar\sigma$ that is much more complex than Eq. (\ref{sigmaeq}). The algebra is left to the reader, but ultimately one obtains a quartic equation for $\bar\sigma$. This is of critical importance, since the fact that we obtained a single possible value for the plateau previously really depended on the fact that Eq. (\ref{sigmaeq}) is of the form $\bar\sigma f(\bar\sigma^2)=0$. It is then obvious that we get a singled valued plateau, since $\bar\sigma=0$ is the necessary trivial solution of no flow and $f(\bar\sigma^2)=0$ gives the symmetric positive and negative solutions required by the orientation symmetry. The BMP model leads to an equation of the form $\bar\sigma f(\bar\sigma^3)=0$. The no flow solution is present, but $f(\bar\sigma^3)=0$ would suggest an asymmetry between positively and negatively defined stresses.\\
Another curious fact is that the BMP model supposes $\xi=0$ and so it is hard to imagine an analytic procedure to obtain the profile $K(y^*)$. And the BMP model makes no mention of the effect of the constraint on the global value of the shear rate, and we have seen that this integral constraint has a key role.

\section{Summary and outlook}
In this article we have adopted an ambivalent tone about the use of the dJS model to rationalize some behaviours of fluids exhibiting shear-banding. This ambivalence is linked to the inherent ambiguity of what we mean when we say that we ``use the dJS model''.\\
\textit{1.} If `using the dJS model' is using it in a purely mechanical approach, in simple shear, it mostly means using the usual syntax of Eqs. (\ref{simpleshear}-\ref{simpleshear2}) and interpreting everything in a strict rheological framework. This approach is the closest to the usual vocabulary used to describe experiments around polymeric material, and therefore it would be known by most of the researchers in the field--`experimentalists' or `theorists'. This approach was essentially underlying most of the recent numerical studies of dJS~\cite{Fielding05,Fielding07,Fielding10}. We outlined this approach in the first three sections (\ref{JSandSketch}-\ref{SimpleShear}). \\
\textit{2.} In earlier studies, in particular by Radulescu \textit{et al.}~\cite{Radulescu99,Radulescu00}, the dimensionless groups that we recall in section~\ref{SimpleShear} were introduced. Those dimensionless groups have two important impacts. Firstly, the dimensionless groups make us realize that the impact of $a$ is identical whatever the precise value of $a$, as long as $|a|\neq 1$. Secondly, the governing equations rewritten for the dimensionless variables allowed Radulescu \textit{et al.} to make very useful connections with general aspects of reaction-diffusion equations~\cite{Radulescu99,Radulescu00}. Radulescu \textit{et al.} `used the dJS model', but in their context it carried a slightly different meaning, influenced by the background on reaction-diffusion equations outlined in section~\ref{ReactDiffInt}. In one of the first studies exploiting the analogy between the dJS governing equations and reaction-diffusion equations, Radulescu \textit{et al.} noted that ``covariant local constitutive equations or microscopic models lead to reaction-diffusion equations with at least two order parameters (shear stress and normal stress difference)''~\cite{Radulescu99}. Actually, there is an exception to this rule, when the stress diffusion only concerns the shear component.\\
\textit{3.} In the case where the stress diffusion only concerns shear stresses, it can be expressed in terms of a diffusion on the symmetric part of the velocity gradient tensor. In this case, Sato \textit{et al.} have shown very recently that the dJS governing equations in simple shear can be synthesised into one single equation for the dimensionless shear rate, Eq. (\ref{kappa_2nd_ODE})~\cite{Sato10}. And they have shown how this equation can be interpreted as representing the motion of a particle in a quartic potential. In this new framework, they are still `using the dJS model'. We have detailed the analytic solution obtained by Sato \textit{et al.} in sections~\ref{PotProp}-\ref{Sato_unconstrained}. We believe that this analytic solution already caries a lot of useful informations. Nevertheless, this solution is inconsistent with the integral constraint given by the value of the global shear rate Eq. (\ref{constraint}), often imposed by the operator in an experimental situation. In particular, we have discussed in section~\ref{leverrulemodif} how this solution becomes increasingly inconsistent as non-local effects become dominant. \\
\textit{4.} To include the integral constraint in a consistent way, we describe in section~\ref{Lagrangiansec} how to use the Lagrangian formalism corresponding to the particle analogy used by Sato \textit{et al.}~\cite{Sato10}. In the Lagrangian formalism, the integral constraint is interpreted as a holonomic constraint. The rigorous solution of the dJS governing equation is the solution to the optimization problem expressed in Eq. (\ref{optimization}). The complete solution will be given elsewhere~\cite{Fardin11}, but we have seen already in sections~\ref{ModFC}~and~\ref{Compet} how the Lagrangian framework allows us to discuss the notion of genuine apparent wall slip and its interaction with the shear-banding state. Following this new trend, we are still `using the dJS model'. \\
\textit{5.} From the Lagrangian formalism of the particle analogy to the dJS governing equations, we just had to make a small step to re-interpret everything in a more `thermodynamic' framework. This was done in section~\ref{RevivalThermo}. The fact that connections exist between the usual `mechanistic/rheological' approach of dJS and a more `thermodynamic' framework was made clear already in the second part of Sato's study~\cite{Sato10}. But the strength of their message was diluted somewhat by the fact that their result was only rigorous close to the critical point $\eta=1/8$. In contrast the free energy we define in Eq. (\ref{freeenergy}) still conveys the full potential of the original framework. We are still `using the dJS model'! Within this framework we attempted in section~\ref{BMPmodBridge} to make connections with the BMP model~\cite{Bautista02,Manero07,Bautista07}. In doing so, we highlighted some points that we hope will stimulate more studies on the connections between the BMP model and more classical rheological frameworks. \\

In summary, we hope that the approaches we detailed here will have given the reader different ``ways of thinking'' about the shear-banding phenomenon. In every approach we are `using the dJS model', but we have been able to go from what could have been thought of as a purely mechanical approach (\textit{1.}), to an approach very reminiscent of statistical field theory and thermodynamics (\textit{5.}). The distinction between the two is obviously less clear cut that was once thought. An interesting point can be made by studying the relative logical implications between the different approaches (\textit{1.}-\textit{5.}). The most general formulation of the law governing the motion of mechanical systems is the principle of least action~\cite{Landau76}, in the sense that the differential equations of motion are implied by the integral formulation. Therefore, the integral thermodynamic approach seems to be more `powerful'. Nonetheless, this approach is linked to the use of an anisotropic diffusion coefficient, except near the critical point~\cite{Sato10}. And more importantly, the form of the potential used in the thermodynamical approach is dependent on the flow geometry. In contrast, the original mechanical approach starts from Eq. ~(\ref{T_gov_diffiso}), which can be evaluated in any geometry. In this respect, the mechanical approach seems to be more powerful. But maybe there exist a systematic procedure to generate the correct potential for any geometry. In particular, as can be derived quickly from Radulescu \textit{et al.}~\cite{Radulescu00}, the spatial inhomogeneity of the stress in a Poiseuille or cylindrical Couette flow~\cite{Larson99,DPL} will translate into a time dependent potential.  We are currently investigating how canonical transformations in the Hamiltonian approach~\cite{Landau76,Goldstein02} can help us introduce new generalized coordinates to solve the problem by making the explicit time dependence of the potential vanish~\cite{Fardin11}. \\
Finally, we have constrained ourselves by using only the dJS model, because it is the most widely used model for shear banding and because, in our opinion, it is one of the simplest tensorial models able to capture shear banding. But the reader should not be constrained by our conservatism. Much more can be done around the dJS model, or around another different basis. For instance, the VCM model is a promising new approach~\cite{Vasquez07}.

\textbf{\normalsize{Acknowledgments }} \\
The authors thanks M. Argentina, S. Asnacios, J.F. Berret, N. Biais, O. Cardoso, A. Colin, S. Fielding, P. Lettinga, S. Manneville, A.N. Morozov, J. Ortin and C. Wagner for fruitful discussions, and the ANR JCJC-0020 for financial support. M.A.F. thanks the Fulbright Commission for their support. T.J.O. acknowledges the NSF Graduate Research Fellowship for funding.


\begin{thebibliography}{50}
\bibitem{homogeneous} Note that in geometries where there exists an inherent inhomogeneity of the stress, this definition will need to be modified to account for the inherent inhomogeneity of the shear rate. This point is connected to the discussion between Adams, Olmsted and Wang [12].
\bibitem{notediv} Actually, this divergence is removed if the constraint on the average value of the shear rate is correctly incorporated. The consistent inclusion of the constraint is discussed in the next sections. Details on how it modifies the interface width will be given elsewhere.
\end{thebibliography}
\end{document}